# On the detection of surface spin freezing in iron oxide nanoparticles and its long-term evolution under ambient oxidation

M Ghoshani[1,2], E H Sánchez[2], S S Lee[3], G Singh[4], N Yaacoub[5], D Peddis[6,7], M Mozaffari[1], C Binns[2], J A De Toro[2] and P S Normile[2]

[1] Department of Physics, Faculty of Science, University of Isfahan, Isfahan 81746-73441, Iran
[2] Instituto Regional de Investigación Científica Aplicada (IRICA) and Departamento de Física Aplicada, Universidad de Castilla-La Mancha, 13071 Ciudad Real, Spain
[3] Institute of Bioengineering and Nanotechnology, 31 Biopolis Way, The Nanos, Singapore 138669, Singapore
[4] School of Aerospace, Mechanical and Mechatronic Engineering, University of Sydney, Sydney, NSW 2008, Australia
[5] Institut des Molécules et Matéríaux du Mans, CNRS UMR-6283, Université du Maine, F-72085 Le Mans, France
[6] Istituto di Struttura della Materia, CNR, 00015 Monterotondo Scalo (RM), Italy
[7] Dipartimento di Chimica e Chimica Industriale (DCCI), Università of Genova, Via Dopdecanes, 16134, Genova, Italy

Email: Peter.Normile@uclm.es and JoseAngel.Toro@uclm.es

## Abstract

Exchange bias effects linked to surface spin freezing (SSF) are commonly found in iron oxide nanoparticles, while signatures of SSF in low-field temperature-dependent magnetization curves have been much less frequently reported. Here, we present magnetic properties of dense assemblies of similar-sized (~ 8 nm diameter) particles synthesized by a magnetite (sample *S1*) and a maghemite (sample *S2*) method, and the influence of long-term (4-year) sample aging under ambient conditions on these properties. The size of the exchange bias field of the different sample (fresh or aged) states is found to correlate with (a) whether a low-temperature hump feature signaling the SSF transition is detected in out-of-phase *ac* susceptibility or zero-field-cooled (ZFC) *dc* magnetization recorded at low field and with (b) the prominence of irreversibility between FC and ZFC curves recorded at high field. Sample *S1* displays a lower magnetization than *S2*, and it is in *S1* where the largest SSF effects are found. These effects are significantly weakened by aging but remain larger than the SSF effects in *S2*, where the influence of aging is considerably smaller. A non-saturating component due to spin disorder in *S1* also weakens with aging, accompanied by, we infer, an increase in the superspin and the radius of the ordered nanoparticle cores. X-ray diffraction and Mössbauer spectroscopy provide indication of maghemite-like stoichiometry in both aged samples as well as thicker disordered particle shells in aged-*S1* relative to aged-*S2* (crystallographically-disordered and spin-disordered according to diffraction and Mössbauer, respectively). The pronounced diminution in SSF effects with aging in *S1* is attributed to a (long-term) transition, caused by ambient oxidation, from magnetite-like to maghemite-like stoichiometry, and a concomitant softening of the spin-disordered shell anisotropy. We assess the impact of this anisotropy on the nature of the blocking of the nanoparticle superspins.

# 1. Introduction

Magnetic nanoparticles (NPs) continue to receive much attention due to a growing number of applications as well as to open questions regarding their rich phenomenology [1–3]. Ferrimagnetic iron oxide NPs with spinel structure (maghemite, $\gamma$-$Fe_2O_3$, and magnetite, $Fe_3O_4$) are particularly interesting due to their biocompatibility, which allows their exploitation in medical applications such as magnetic hyperthermia, contrast agents for magnetic resonance imaging and targeted drug delivery [1,4–7]. An important feature of iron oxide-based NPs is the frequent appearance of surface spin disorder (SSD) or canting, which can be intrinsic, caused by broken exchange bonds at particle surfaces [8–12], or due to surface structural disorder [13]. With decreasing particle size, SSD often becomes more prevalent [14–20] but can also be suppressed due to improvement in (intraparticle) crystalline order [13]. SSD can also become more noteworthy upon moving to a hollow particle morphology [21,22]. At ambient temperature, SSD can behave as a magnetically dead layer, reducing the overall saturation magnetization [9,10,18,23]. Upon cooling such layers can undergo spin glass-like freezing [24,25]. Due to the high magnetic anisotropy of such surface spin glass-like phase relative to the magnetically ordered NP cores, an exchange bias (EB) effect – i.e. the displacement of a low-temperature hysteresis loop (recorded after field cooling) along its applied field axis – is often observed in magnetic NPs possessing SSD [26].

    A variety of other magnetic features have also been related to SSD. These include open hysteresis loops and large differential susceptibility at high fields [20,24,27], irreversibility between field-cooled (FC) and zero-field-cooled (ZFC) *dc* magnetization curves measured at high field [10,24,28], and an increase in the saturation magnetization (relative to an extrapolation based on Bloch's law) below the surface spin freezing (SSF) temperature [25,29]. However, in contrast to EB effects, such features are *not* observed systematically across the vast literature of magnetic oxide NPs with SSD. In particular, the direct detection of SSF via the appearance of a hump (or minor peak compared to that due to particle blocking) in low-field ZFC *dc* magnetization or in *ac* susceptibility curves has only rarely been reported [30–33]. The latter technique (*ac* magnetometry) probes spin dynamics and therefore is capable of distinguishing SSF from conventional particle blocking [11,13,30–32,34–37].

    In the present article we review the factors leading to fingerprints of SSF in the thermal dependence of *ac* and *dc* magnetization, studying iron oxide NP samples that (as will be concluded) differ essentially with respect to aspects of their spin disordered shells. We correlate the size of such fingerprints with other signatures of SSD (in particular, the EB field but also enhanced coercivity, reduced magnetization values, and high-field FC-ZFC irreversibility). The sensitivity of these techniques in the detection of SSF is explored through two types of comparison between ferrimagnetic iron oxide NPs of similar size distribution: (i) the comparison of samples prepared by two different synthesis methods, a magnetite

(sample *S1*) and a maghemite (sample *S2*) method, and (ii) the study of long-term aging (oxidation under ambient conditions) on the SSF effects in the samples.

## 2. Experimental methods

Sample *S1* was prepared by a procedure based on the thermal decomposition of iron oleate in the presence of oleic acid and hexadecane. This procedure was previously reported to produce magnetite ($Fe_3O_4$) NPs [38,39]. Sample *S2* was prepared by employing a method that had previously been reported to yield maghemite ($\gamma$-$Fe_2O_3$) NPs – namely a procedure based on thermal decomposition of $Fe(CO)_5$ in the presence of an oleic acid and dioctyl ether, followed by oxidation with trimethylamine N-oxide [13]. (Additional information on NP synthesis, particularly on *S1*, is provided in the supplementary material.) Both *S1* and *S2* particles were precipitated by adding acetone to their solutions. The particles were then collected by centrifugation to yield powders of oleic acid (OA) coated NPs. The OA contents by mass were determined by thermogravimetric analysis to be, approximately, 10 % in *S1* and 20 % in *S2*. Disk-like samples of these NP powders for use in magnetic measurements were prepared by die pressing under approximately 0.7 GPa. These *S1* and *S2* samples constitute dense NP assemblies: the magnetic NP packing fraction ($\phi$) is assumed to be around 0.5 in *S2* (the value previously obtained in a pressed disk-like sample of similar OA-coated maghemite NPs [40]) and is expected to be a little higher in *S1* (around 0.55 – see section VIII of the supplementary material) due to its lower OA content relative to *S2*.

Transmission electron microscopy (TEM) images were obtained using a FEI Tecnai G2 F20 microscope operated at 200 kV. Magnetic measurements were performed using a Quantum Design MPMS Evercool *SQUID* (superconducting quantum interference device) magnetometer. Hysteresis loops were obtained at different (low) temperatures after first performing sample cooling from room temperature to 5 K in a field of 50 kOe applied parallel to the sample (disk) plane. Temperature dependent *dc* and *ac* magnetization curves were measured upon heating from 5 K after either the ZFC (in *ac* and *dc* measurements) or FC (*dc* only) state had been prepared. The applied field (and cooling field in FC measurements) was 5 Oe in "low-field" *dc* measurements, 50 kOe in "high-field" *dc* measurements, and oscillatory of frequency 10 Hz and amplitude of 2.5 Oe in *ac* measurements. X-ray diffraction (XRD) measurements were carried out in transmission geometry (using a few milligrams of NP powder mounted on Kapton tape), employing a Bruker D8 diffractometer operating with Cu K$\alpha$ radiation and a position sensitive (silicon strip) detector. $^{57}$Fe Mössbauer spectra were obtained at 10 K using an applied field of 80 kOe oriented parallel to the incident $\gamma$-beam (obtained from a $^{57}$Co/Rh $\gamma$-ray source mounted on an electromagnetic transducer controlled according to a triangular velocity waveform).

All *SQUID* measurements were performed both on "fresh" NPs (particles that had been exposed to ambient conditions for no more than several weeks since being synthesized) and "aged" NPs (particles, from the same sample batches, that had been stored under ambient conditions for a 4-year period), except for the high-field (temperature dependent) magnetization curves, which were obtained only in aged NP samples. TEM images were taken on fresh NPs. XRD and Mössbauer spectroscopy were conducted on aged NPs. In what follows, *S1* and *S2* (*S1a* and *S2a*) refer to the fresh (aged) sample states.

## 3. Results and discussion

Figure 1 shows TEM images of *S1* and *S2* NPs. In both cases the NPs are found to be approximately spherical and to correspond to a similarly narrow size distribution ($\sigma \approx 2$ % from log-normal fits of the size histograms), of average diameter ($D_{TEM}$) of 8.6 nm in *S1* and 8.0 nm in *S2*. High resolution TEM suggests a higher degree of crystalline order in *S2* than in *S1*.

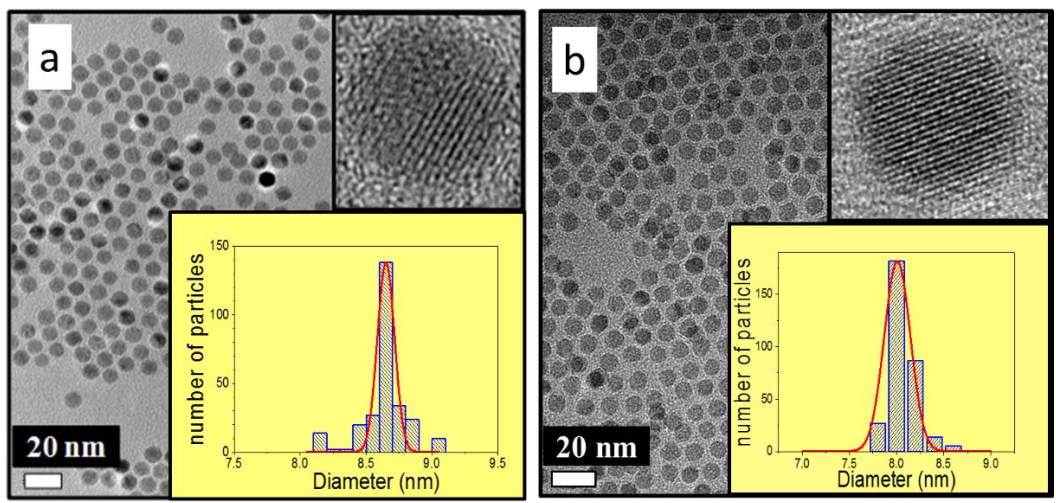

**Figure 1**. Transmission electron micrographs of (a) *S1* and (b) *S2* nanoparticles. The upper insets are representative high-resolution TEM images of the particles. Each lower inset shows the particle size distribution (extracted from TEM images) fitted to a log-normal function.

Figure 2 presents low-temperature hysteresis loops of *S1*, *S1a* and *S2*. The loop of *S2a* is very similar to that of *S2* and therefore (for clarity in presentation) has been omitted from figure 2 (the *S2a* loop is provided in figure S1 of the supplementary material). The (mass) magnetization (M) signals of the different samples have been corrected for OA mass fraction (see supplementary material for details). Prior to performing these magnetic measurements, we expected *S1* would have a larger magnetization than *S2* owing to the NP synthesis methods employed (*S1*: magnetite method; *S2*: maghemite method), the saturation magnetization ($M_S$) of bulk magnetite being around 16% larger than that of bulk maghemite.

Remarkably, however, we observe (in figure 2) that $M_S$ of *S1* appears to be considerably smaller than that of *S2*, which suggests a more appreciable fraction of SSD ("missing magnetization") in *S1* NPs as compared to *S2* NPs. We use the word "appears" due to the presence of the non-saturating (up to the maximum applied field, $H_{max}$ = 50 kOe) component in the *S1* loop. Hereafter we refer to the "technical $M_S$" (i.e. the magnetization value corresponding to $H_{max}$) when discussing this sample.

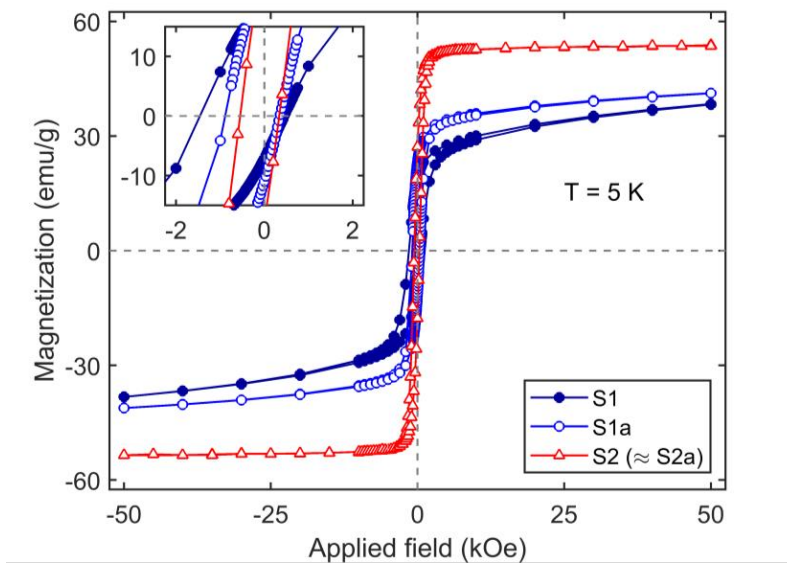

**Figure 2**. Hysteresis loops recorded at a temperature of 5 K in the *S1* sample, when fresh and aged (*S1a*), and in the *S2* sample when fresh (a very similar loop is found in aged-*S2* – see supplementary material). The inset is a close-up indicating where each branch (of each loop) intercepts the applied field axis.

At room temperature, the technical $M_S$ of *S1* is (as expected) around 10 % lower than at 5 K and the M(H) dependence in *S1* continues to exhibit a non-saturating component (see figure S3 of the supplementary material, *after the References section*). In contrast to the magnetic response of *S1*, near-bulk magnetite $M_S$ values (60-80 emu/g) were found in a recent study of magnetite NPs performed by Kemp *et al*. [41]. In that study, the average particle volume of the NPs was an order of magnitude larger than that of the *S1* particle, and a non-saturating component was *not* detected in room temperature M(H) measurements. The presence of the non-saturating component in *S1* is therefore linked to its greatly reduced $M_S$ value with respect to bulk magnetite (in subsection 3.1, we attribute it the progressive alignment of spins from disordered particle shells). The $M_S$ value in *S2* (figure 2) is lower than the bulk maghemite room temperature value (~74 emu/g) but by a lesser extent than the (apparent) discrepancy between the $M_S$ values of *S1* and bulk magnetite.

A clear EB effect (displacement from H = 0 of each hysteresis loop center) is observed in each loop of figure 2 (see the figure inset). The effect is attributed, in each sample and (fresh or aged) state, to exchange coupling between a ferrimagnetically ordered iron oxide NP core and a magnetically disordered (and more anisotropic) iron oxide shell. Values of EB field ($H_E$) and coercive field ($H_C$) extracted from these loops are presented in table 1. The correlation between $H_E$ and $H_C$ indicates that coercivity enhancement (by exchange coupling) [42,43] is significant in the samples. Compared to the range of $H_E$ values reported in the literature on magnetite NPs of similar size to those presented here – which is rather wide and includes the case of *no* EB effect [14,15,44,45] – the value measured in *S1* is relatively large [27]. The $H_E$ value in *S2*, on the other hand, is typical of maghemite NPs [10,18,46,47]. Our findings are consistent with the presence of relatively thicker (as already suggested above based on the discussion of $M_S$ values) and more anisotropic shells of SSD in *S1* NPs relative to *S2* NPs. The aging effect on the EB properties and the non-saturating component of *S1* will be discussed towards the end of this section (in subsection 3.1), where we will also discuss the inference (from figure 2) of an increase in the *S1* NP core superspin with aging.

In figure 3 we explore the correlation between the temperature dependence of the EB effect – figure 3(d) – and the presence (or absence) of various temperature-dependent SSF-related features in the out-of-phase *ac* susceptibility ($\chi''$) – figure 3(a) – and low-field ZFC magnetization curves ($M_{ZFC;5Oe}$) – figure 3(c). Each main peak in $\chi''$ and $M_{ZFC;5Oe}$ – situated between 90 and 110 K in $\chi''$ and between 103 and 116 K in $M_{ZFC;5Oe}$ – is due to blocking of NP moments (*superspins*) that is influenced or governed by interparticle magnetic dipolar interactions. Such interactions – which have been extensively studied by ourselves and others in similar NP assemblies, particularly in relation to superspin glass behaviour [37,48–51] – do *not* constitute the main interest in the present study. However, at the end of this section (namely subsection 3.2) we will discuss why we believe the nature of the blocking in *S1* is governed by a combination of interparticle interactions and a relatively high single-particle energy barrier due to the significant SSD in the *S1* NPs. (In subsection 3.2 we will also account for the difference in the shapes of the main peaks in the $\chi''$ curves of *S1* and S2.)

A low temperature hump feature is clearly exhibited in the $\chi''$ curve of *S1*. In figure 3(b) this curve is shown fitted to a sum of two gaussian functions, one representing the NP blocking transition and the other (the weaker gaussian) representing the hump feature. The position of maximum negative slope of the smaller gaussian fitting curve of figure 3(b) is located around 65 K, which coincides with the position of the onset of EB ($T_E$) in *S1* – figure 3(d). In *ac* susceptibility studies of canonical spin glasses, the freezing temperature (at a given measurement frequency) is usually determined as either the temperature position of the peak in the in-phase susceptibility ($\chi'$) or of the inflection in $\chi''$ [52]. It is therefore reasonable to attribute the hump feature in the $\chi''$ curve of *S1* to an SSF transition at $T_{SSF} \approx 65$ K in the *S1* NPs. We are aware of just one previous study in which a low temperature peak in a $\chi''$ curve was associated with SSF in iron oxide-based NPs (the particles being nickel ferrite) [30].

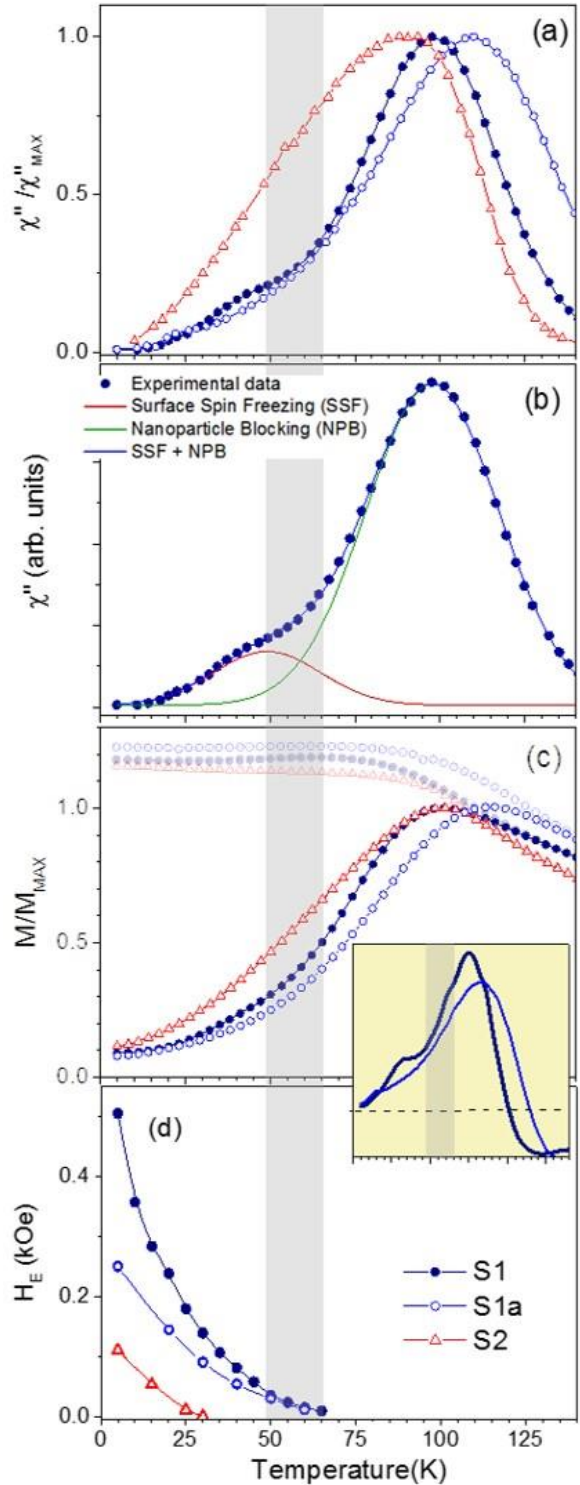

**Figure 3**. Temperature dependence of (a) out-of-phase ($\chi''$) *ac* susceptibility, (c) low-field FC and ZFC *dc* magnetization and (d) exchange bias field of the *S1* (fresh and aged) and *S2* (fresh) samples. (b) Fit to the $\chi''$ curve of fresh-*S1* using two gaussians. The rectangular shaded region indicates the temperature range in which SSF take places in *S1* (fresh and aged). Inset of (c): the derivative with respect to temperature of the ZFC magnetization of fresh-*S1* (dark blue curve) and aged-*S1* (lighter blue curve).

The presence of an albeit less conspicuous SSF hump feature contributing to the $M_{ZFC;5Oe}$ curve of *S1* – figure 3(c) – is suggested by the shape of that curve just below $T_E$ of *S1*. This possibility is supported by the shape of the derivative with respect to temperature of that curve – plotted in the inset of figure 3(c) – where a slight minimum is detected around 50 K. That a signature of SSF in a low-field *dc* magnetization curve should occur at a lower temperature than in an *ac* measurement is indeed expected from the typical behaviour of the freezing temperature with measurement frequency in a spin glass-like phase [30,52]. It is therefore plausible to conclude the presence of an SSF hump signature in the low-field ZFC curve of *S1*. Such signatures have only rarely been reported in oxide-based NPs [31-33].

With both samples (*S1* and *S2*) in their aged states, it was decided to investigate a further effect related to SSF in magnetic NPs – namely high field FC-ZFC irreversibility [10,24,28]. We observe – figure 4 – that *S1a* exhibits considerably larger FC-ZFC irreversibility than *S2a*, which is reasonable since aged-*S1* continues to exhibit considerably larger EB parameters ($H_E$ and $H_C$ values in table 1) than both fresh and aged *S2*. The bifurcation temperature corresponding to each irreversibility effect is consistent with the EB onset temperature of each sample ($T_E$ was not measured in *S2a* but is assumed to be the same as in fresh-*S2*).

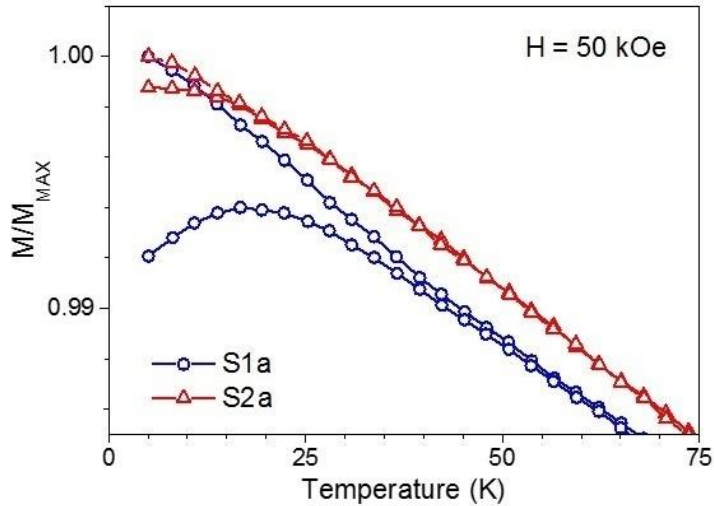

**Figure 4**. FC and ZFC *dc* magnetization curves measured at high field in both aged samples.

As summarized in table 1, the presence (detectability) of the SSF features in the *ac* and *dc* (low-field ZFC) magnetization curves of figure 3 correlates with the samples' EB properties. Namely, the features are best seen in the sample with the largest $H_E$ (*S1*), are still (barely) visible in that same sample after aging (*S1a*), where an intermediate $H_E$ value is found, and are absent in the sample with the lowest $H_E$ (*S2*). As mentioned, the observation of such a clear SSF fingerprint in the temperature dependence of (*dc* or *ac*) magnetization is rather unusual; we are aware of just a few studies reporting such features in the large body of literature devoted to SSD in magnetic oxide particles [30–33], with just one of these studies dealing with NPs of a pure iron oxide (magnetite) [33]. This would suggest that for the detection of such fingerprints, iron oxide NPs require disordered shells of a relatively large thickness and high anisotropy not found in the majority of studies reported to date. The size of the FC-ZFC irreversibility (quantified in table 1) also correlates with EB properties for case of the two (aged) sample states.

**Table 1.** Summary of SSF-related magnetic results in the *S1* and *S2* samples in their fresh and 4-year aged ("*a*") states: EB field ($H_E$) and exchange-enhanced coercivity ($H_C$) at T = 5 K from figure 2 and figure S1 of supplementary material; EB onset temperature ($T_E$) from figure 3(d); brief description of any SSF-related feature in the temperature dependence of the imaginary component of *ac* susceptibility ($\chi''$) or the ZFC *dc* magnetization at low applied field ($M_{ZFC;5Oe}$) in figure 3 (a)-(c); and a quantification of the degree of irreversibility between FC and ZFC high-field magnetization curves, $((M_{FC} - M_{ZFC})/M_{FC}) \times 100$ detected at {T,H} = {5 K, 50 kOe}, in figure 4. The penultimate column shows the temperature position ($T_{max}$) of the main peak in each $M_{ZFC;5Oe}(T)$ curve of figure 3(c) and figure S2 of supplementary material. We conclude (main text) that $T_{max}$, as well as being determined by interparticle dipolar interactions, is affected by high single-particle anisotropy (due to SSD) in *S1*. The final column gives the ratio between the estimated size of the NPs' crystallographically-ordered cores ($D_{Scherrer}$), obtained from XRD on aged NPs (figure 5), and the average (geometric) particle size ($D_{TEM}$), obtained from TEM on the fresh NPs (figure 1). *NA* (not available) indicates that the given parameter was *not* measured for the given (fresh or aged) sample state.

| *Sample* (*fresh or aged*) | $H_E$ (Oe) | $H_C$ (Oe) | $T_E$ (K) | *SSF feature in ac or low-field dc T-dependent magnetization curve* | $\left(\frac{M_{FC} - M_{ZFC}}{M_{FC}}\right) \times 100$ | $T_{max}$ (K) | $\frac{D_{Scherrer}}{D_{TEM}}$ |
|---|---|---|---|---|---|---|---|
| *S1* | 505 | 953 | 65 | *Clear* hump in $\chi''(T)$ at ~ 65 K (= $T_E$). *Weak* hump in $M_{ZFC;5Oe}(T)$ at ~ 50 K. | NA | 103 | NA |
| *S1a* | 267 | 620 | 65 | *Weak* hump in $\chi''(T)$ at ~ 65 K. Hump *no* longer in $M_{ZFC;5Oe}(T)$. | 0.80 | 116 | 0.76 |
| *S2* | 106 | 442 | 30 | *No* SSF feature in $\chi''(T)$ or $M_{ZFC;5Oe}(T)$. | NA | 103 | NA |
| *S2a* | 95 | 346 | NA | *No* SSF feature in $\chi''(T)$ or $M_{ZFC;5Oe}(T)$. | 0.14 | 103 | 0.83 |

To gain insight into how the *S1* and *S2* NPs differ with respect to disordered shell content as well as stoichiometry, XRD and Mössbauer spectroscopy were carried out on aged samples of these NPs. The XRD patterns – figure 5 – only contain Bragg reflections corresponding to a spinel ferrite structure. Using the widths of all the reflections of each pattern, we estimate – see section V of the supplementary material – similar values of Scherrer grain size ($D_{Scherrer}$), namely 6.5 nm (*S1a*) and 6.6 nm (*S2a*). Bearing in mind that TEM (figure 1) indicates a slightly larger geometrical particle size for *S1* ($D_{TEM}$ = 8.6 nm) compared to *S2* ($D_{TEM}$ = 8.0 nm), these grain size estimates imply a larger crystallographically-disordered volume fraction in the *S1a* NPs relative to *S2a* NPs (i.e. the ratio $D_{Scherrer}/D_{TEM}$ is smaller for *S1a*, implying a larger disordered volume fraction). Assuming that thicker spin disordered shells are a concomitant of having thicker crystallographically-disordered shells, these results are consistent with our expectation of a more appreciable fraction of SSD in the *S1* NPs relative to the *S2* particles. The extracted lattice parameter (inset of figure 5) is slightly closer to the value of bulk maghemite in *S2a* than in *S1a*. Taken in isolation, of course, this does not allow us to address the question of the stoichiometry (maghemite or magnetite) of the NPs [53,54]. The following Mössbauer analysis helps in this respect.

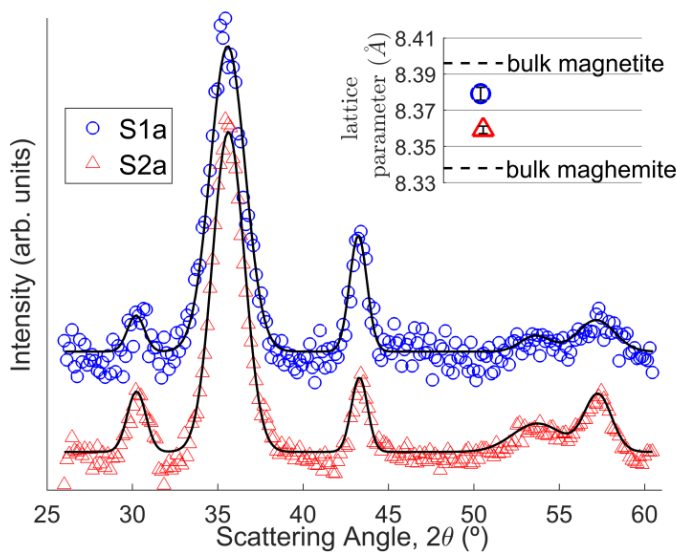

**Figure 5**. X-ray diffraction from aged nanoparticles of *S1* and *S2*. The solid lines are fits using a profile-matching technique. The inset shows the comparison of the extracted lattice parameters with those of bulk maghemite and magnetite.

Mössbauer spectra of *S1a* and *S2a* are presented in figure 6. Applied to iron oxide-based materials in a large external magnetic field (as employed here), this spectroscopic technique provides accurate information on the oxidation states and cationic distribution of different Fe species, and it allows Fe ions located at tetrahedral (A) and octahedral (B) interstitial sites to be distinguished. Detailed analysis of the spectra of figure 6 – the fitting parameters of which are presented in table 2 – indicates that *S1a* differs from natural magnetite (where $Fe^{2+}$ ions would be present and characterized by a large isomer shift and a small hyperfine field, and the B site contribution to the spectrum would be complex and asymmetrical [55]). Namely we find that the isomer shift and hyperfine field values in *S1a* (as well as in *S2a*) are more typical of maghemite NPs [56]. Based on this result together with the lattice parameter analysis, we suggest that the effect of long-term aging on the SSF phenomena in *S1* is related to a (long-term) transition – caused by ambient oxidation – towards a maghemite-like stoichiometry of NPs (*S1*) that were initially magnetite-like [57]. This idea is developed further in the following subsection.

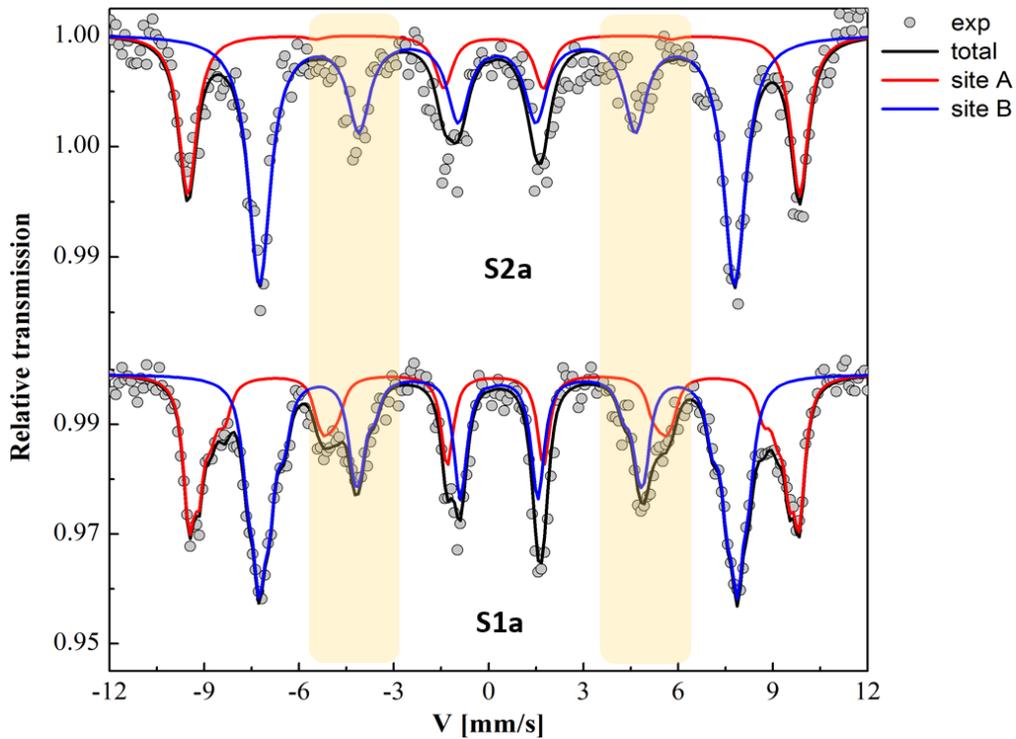

**Figure 6**. Mössbauer spectra obtained at 12 K and H = 80 kOe in aged nanoparticles of *S1* and *S2*. The solid black (red or blue) lines are total (partial) fits to the data.

In *S2a* there exists a significant value of $Fe_B^{3+}/Fe_{total}^{3+}$ at the expense of the value of $Fe_A^{3+}/Fe_{total}^{3+}$ (table 2), suggesting an excess of octahedral Fe sites at the surface of maghemite NPs. Crucially, both high-field spectra (figure 6) exhibit second and fifth lines of non-zero intensity, which usually indicates a canted structure for $Fe^{3+}$ magnetic moments, with respect to the applied field, i.e. a non-collinear magnetic structure [8,47,58,59]. In *S2a* we observe a significant spin-canting only at the B site (41°, table 2), which may be related to the aforementioned excess of octahedral sites at the NP surface. In contrast, both A and B sites present a strong spin-canting in *S1a*. Assuming a simple ordered-core/disordered-shell magnetic morphology in which a spin-disordered shell accounts for all the observed canting, these results support the idea of a relatively thicker spin-disordered shell in *S1a* compared to *S2a*.

**Table 2.** Parameters obtained from the fits of the Mössbauer spectra in figure 6: isomer shift ($\delta$), quadrupole shift ($2\varepsilon$), effective field ($B_{eff}$), hyperfine field ($B_{hyp}$), average canting angle ($\theta$) and ratio of the A and B sites. The isomer shift values are referenced relative to $\alpha$-Fe at 300 K. The values in parentheses are uncertainties and refer to the least significant digit. The estimated error on each $\theta$ value is $\pm 10°$.

| Sample | Site | $\delta$ (mm s$^{-1}$) | $2\varepsilon$ (mm s$^{-1}$) | $B_{eff}$ (T) | $B_{hyp}$ (T) | $\theta$ (°) | $Fe^{3+}_{A,B}/Fe^{3+}_{total}$ |
|---|---|---|---|---|---|---|---|
| *S2a* | $Fe^{3+}_A$ | 0.33(1) | 0.00(1) | 59.8(2) | 52.0(2) | 11 | 0.29(1) |
|  | $Fe^{3+}_B$ | 0.45(1) | 0.00(1) | 46.3(2) | 52.6(2) | 41 | 0.71(1) |
| *S1a* | $Fe^{3+}_A$ | 0.38(1) | -0.04(1) | 57.3(2) | 51.7(2) | 42 | 0.41(1) |
|  | $Fe^{3+}_B$ | 0.51(1) | -0.03(1) | 46.2(2) | 52.2(2) | 43 | 0.59(1) |

### 3.1. *Ordered-core volume fractions and the effect of aging in S1 on NP core size and disordered shell anisotropy*

The non-saturating components in figure 2, although *not* associated directly with the SSF transition at $T_{SSF} \approx 65$ K (since such a component is observed far above $T_{SSF}$ in the M(H) curves from *S1* in figure S3 of the supplementary material), are understood to due to the progressive alignment of spins from the disordered NP shells. This understanding is based on previous work of iron oxide-based NPs, where such components were observed (at low and room temperature) in hollow maghemite NPs and were attributed to SSD [46], and where, in a very recent study of cobalt ferrite NPs, it was demonstrated that such components are due the progressive alignment of spins from otherwise spin-disordered shells (alignment that extends radially outwards from the interface of each ferrimagnetically ordered core, thus increasing the effective size of that core with field) [60]. The (mass) magnetization value at the point around which each hysteresis loop – of figure 2 – closes provides an estimate of the *superspin* ($\mu$) of each ferrimagnetically ordered NP core ("uncontaminated by a magnetic moment signal from SSD") divided by the total mass (ordered and disordered regions) of the NP. The extraction of these values (denoted as $M_{cores}$) is indicated in section VI of the supplementary material. Dividing the $M_{cores}$ values of *S1a* and *S2a* by a typical value of magnetization of bulk maghemite at 5 K (~ 80 emu/g) provides estimates for the magnetically ordered volume fraction of each NP type – *viz.* 0.44 (*S1a*) and 0.66 (*S2a*) – which are roughly consistent with volume fractions of crystallographically-ordered phase obtained from XRD; namely $(D_{Scherrer}/D_{TEM})^3$ values of 0.43 (*S1a*) and 0.56 (*S2a*).

The reduction in the gradient of the non-saturating component with aging (figure 2) appears to be accompanied by an increase in $\mu$ (in other words, a slight *reduction* in "missing magnetization" with aging), which would suggest – since we would not expect an increase in the "microscopic volume magnetization" ($\mu$ divided by NP core volume) under long-term ambient oxidation – an increase in the radius of the ordered NP core. Using $M_{cores}$ values for *S1* and *S1a* together with the $D_{Scherrer}$ value for *S1a*, we estimate (see section VII of the supplementary material) that the radius of the ferrimagnetically ordered core increases with aging by $\Delta r_{core}$ of between 0.13 and 0.34 nm, which would correspond to the interface between the ordered core and disordered shell "swelling out" to incorporate (approximately) an additional monolayer of iron oxide from the shell into the core.

Aging of *S1* yields a significant reduction in the EB properties but it is not found – see figure 3(d) – to affect $T_E$, which is consistent with the appearance of the $\chi''(T)$ curves of this sample – figure 3(a) – which each possess a low temperature hump (albeit a weak one in *S1a*) indicative of essentially the same $T_{SSF}$ value ($\approx$ 65 K). Considering $T_{SSF}$ as a property analogous to the ordering temperature ($T_N$) of the high anisotropy (antiferromagnet) component in conventional EB systems [42,43], and bearing in mind that $T_N$ (which represents the upper limit value of $T_E$ in such systems) has been reported to decrease strongly upon moving from nano to subnanometric layer thickness [61,62], the lack of variation in $T_E$ (or, equivalently, $T_{SSF}$) from *S1* to *S1a* would suggest that the thickness of the spin-disordered shells is not significantly altered by aging. This is consistent with the at most 0.34 nm growth (reduction) in the NP core radius (in the disordered shell thickness) estimated above. We suggest that it is a softening (decrease) in the anisotropy of the SSD caused by aging in *S1* which is the decisive factor accounting for the diminution in the SSF (including EB) effects. Since the magnetic anisotropy is lower in the ferrimagnetic phase of bulk maghemite than in that of bulk magnetite, we suggest that such softening is the result of the gradual transition towards maghemite-like stoichiometry caused by long-term ambient oxidation in *S1*.

## 3.2. Impact of surface spin disorder on the blocking of nanoparticle superspins

Finally, we discuss the impact of the SSD on the nature of the blocking of the nanoparticle superspins. Both *S1* and *S2* samples exhibit the same value of main peak position ($T_{max}$ = 103 K) in their $M_{ZFC;5Oe}$ curves – figure 3(c). Upon first consideration, this may be interpreted (mistakenly) as an indication of identical values of the interparticle dipolar interaction strength parameter, $E_{dd}$ ($\propto \mu^2/d^3$, where $d$ is the mean center-to-center separation between nearest-neighbour NPs in each assembly and $\mu$ is the NP core superspin, defined already), for *S1* and *S2*. However, by more careful consideration (section VIII of the supplementary material) we estimate that the ratio $E_{dd;S1}/E_{dd;S2}$ is at most only around 0.5. This implies a significant contribution to $T_{max}$ arising due to a relatively large effective single-particle magnetic anisotropy ($K_{eff}$), in turn due to a large and highly-anisotropic SSD fraction in *S1*. The variation in $T_{max}$ with aging in *S1*, namely the value of the ratio $T_{max;S1}/T_{max;S1a}$ (= 0.89 – see table 1), cannot be explained alone by our estimated value of the ratio $E_{dd;S1}/E_{dd;S1a}$, our upper value of which is 0.79 (see section VIII of supplementary material). This also implies a contribution to $T_{max}$ due to $K_{eff}$, namely a contribution that weakens with aging ($K_{eff}$ weakens with aging in *S1* and hence $K_{eff;S1}/K_{eff;S1a} > 1$) in line with the sample's aged-induced reduction in SSF effects. Very recently it was suggested, by some of use, that individual (single-particle) magnetic anisotropy impacts on dipolar collective properties of dense assemblies of cobalt-ferrite NPs [63].

As pointed out some time ago by Mørup [49], the interparticle interaction strength in a dense magnetic NP system should be quantified relative to the single-particle anisotropy energy barrier ($K_{eff}V_{NP}$, where $V_{NP}$ denotes average NP volume). Under such normalization, *S2* is yet a stronger interacting system (and hence is yet more likely to *super*spin glass-like) than *S1*, i.e. as well as *S2* having an approximate factor of two higher $E_{dd}$ parameter than *S1*, the *S2* NPs will have a lower $K_{eff}$ value than *S1*, as a consequence of possessing a lower and less anisotropic SSD fraction. The $E_{dd}/(K_{eff}V_{NP})$ ratio for *S2* is therefore expected to be more than a factor of two larger than the $E_{dd}/(K_{eff}V_{NP})$ ratio of *S1*. Our expectation is that *S1* is only a moderately interacting system (close to the border of becoming superspin glass) while *S2* is more strongly interacting (and hence superspin glass-like). This expectation is confirmed by additional *SQUID* measurements presented in figures S4 and S5 of the supplementary material.

The difference in $E_{dd}/(K_{eff}V_{NP})$ ratio (in part due to the difference in $K_{eff}$ values) between *S1* and *S2* is manifest in the difference in shapes of the main peaks in their $\chi''$ curves – figure 3(a). The broad shoulder on the low temperature side of the peak position in the *S2* ($\chi''$) curve is attributed to the combination of an asymmetrical line-shape intrinsic to the physics of superspin glasses (a steeply rising curve – i.e. sudden onset of $\chi''$ – on the high temperature side of the peak, and a broad tail on the low temperature side), and demagnetizing field (DMF) effects, which – as was established in a recent study on similarly dense assemblies of maghemite NPs [40] – tend to accentuate such asymmetry. The main peak in the *S1* ($\chi''$) curve, in contrast, possesses no such broad shoulder and can be fitted – figure 3(b) – to a symmetrical (gaussian) function, a consequence of the lower $E_{dd}/(K_{eff}V_{NP})$ ratio in *S1* relative to *S2*. (See section X of the supplementary material for notes on the DMF factors and for $\chi''(T)$ data – figure S6 – measured on a pressed disk-like sample of NPs from the same batch as *S2* but coated with silica shells, instead of oleic acid. Figure S6 simultaneously serves to (i) highlight the loss of the intrinsic asymmetrical line-shape as interparticle interactions are reduced relative *S2* and (ii) rule out the possibility that DMF effects prevented a hump signature of SSF from being detected in *S2*. See the text below that figure for more details.)

# 4. Conclusions

Prior to the present study, several different magnetic effects had been reported as being related to surface spin disorder in a large body of literature on iron-oxide based NPs [8-37,46-47,58-60]. By reviewing some of these effects in two samples of similar NP diameter (approximately 8 nm) but synthesized by different methods (one for magnetite and the other maghemite), we have been able to determine why SSF-related hump features in out-of-phase *ac* susceptibility and low-field *dc* (ZFC) magnetization curves had been seldom seen in pure iron oxide NPs. Namely, we have found that the requirement for the clear appearance of such features is a relatively thick spin-disordered shell (a shell volume of more than 50 % of total NP volume) of a relatively high anisotropy (pertaining more to spin-disorder in magnetite than maghemite). These conditions also give rise to a strong (for iron oxide NPs) exchange-bias effect (EB field, $H_E \sim 500$ Oe, the case of sample *S1*, produced by the magnetite method). When the conditions are not fulfilled (the case of sample *S2,* produced by the maghemite method), an EB effect ($H_E \sim 100$ Oe) can still comfortably be detected, suggesting that the measurement of EB field may provide the most sensitive fingerprint of SSF in iron oxide NPs. The SSF effects in the sample that initially fulfills those conditions (sample *S1*) have been found to become strongly degraded by long-term (4-year) aging under ambient conditions, which, supported by XRD and Mössbauer analysis, we have attributed to oxidation towards a maghemite-like stoichiometry that results in a softening of the magnetic anisotropy associated with the spin-disordered shells. Finally, we have been able to connect aspects relating to the blocking of NP superspins (in *S1*, *S2* and aged-*S1*) to expected differences in effective single-particle anisotropy (largely governed by surface spin disorder) between samples. In future work it would be interesting to study a dense assembly of NPs similar to *S1* as a function of progressive annealing in air, at moderate temperatures (below the oleic acid boiling point), assessing changes in SSF effects and correlating them with structural information from small angle neutron [60] and total (x-ray) scattering [64].

**Acknowledgments**

This work was supported by the Spanish Ministerio de Economía y Competitividad (grant number MAT2015-65295-R). Maral Ghoshani gratefully acknowledges travel funding from the Office of Graduate Studies, University of Isfahan. We thank Dr Rosario Ballesteros for performing thermogravimetric analysis.

**ORCID iDs of corresponding authors**

PSN: https://orcid.org/0000-0002-8851-9899

JADT: https://orcid.org/0000-0002-9075-1697

# Supplementary Material

Contents

# I. Additional information on nanoparticle synthesis

The detailed procedure to synthesize the *S1* nanoparticles was as follows. First iron oleate was prepared by mixing 20 mmol of iron (III) chloride ($FeCl_3 \cdot 6H_2O$, 98 %) and sodium oleate (60 mmol) in a 250 mL round-bottom flask containing deionized water (30 mL), hexane (70 mL) and ethanol (40 mL). The reaction mixture was then stirred under ambient atmosphere for 4 hours at 70 °C. The dark red organic product was separated from the water phase and washed three times with deionized water to discard reaction by-products. The waxy solid product (iron oleate) was obtained after evaporation of hexane and residual water content by using a Heidolph rotary evaporator (at a bath temperature of 70 °C). To synthesize spherical nanoparticles, the prepared iron oleate (1.62 g, 1.8 mmol) and oleic acid (0.6 mL, 1.9 mmol) were added to a 100 mL round-bottom flask containing 25 mL of hexadecene. The reaction mixture was heated to 300 °C at the rate of 3 °C/min under argon atmosphere and then refluxed at this temperature for 45 min before being cooled down to 100 °C. The reaction product was washed three times with toluene and isopropanol. Regarding the preparation of sample *S2*, the size of the final nanoparticles was controlled by carefully adjusting the amount of oleic acid. The reaction temperature was precisely controlled by a temperature controller with a thermocouple inserted in the reaction solution.

# II. Low T hysteresis loop and ZFC curve of *S2a* compared to *S2*

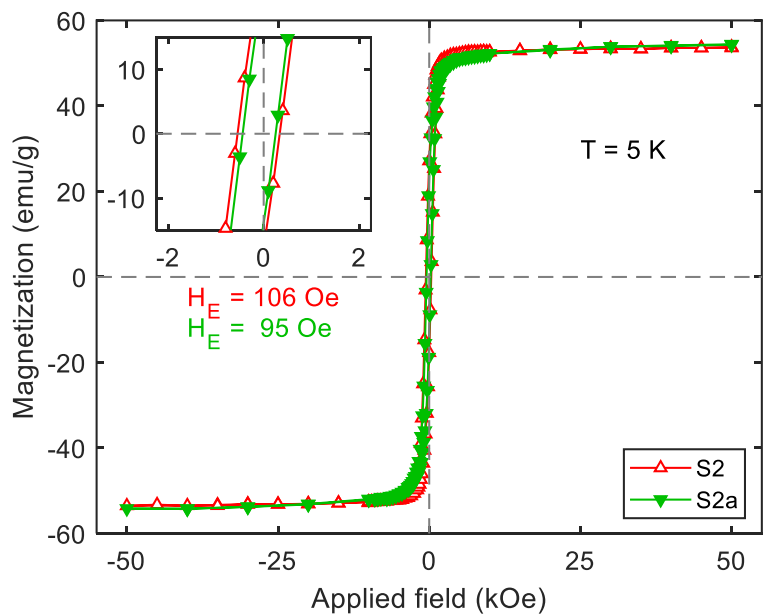

**Figure S1. Low temperature hysteresis loops of the fresh- and aged-*S2* sample**
The inset is a close-up indicating where each branch intercepts the field axis. Both loops were recorded after sample cooling from room temperature to 5 K in the presence of an applied field of 50 kOe.

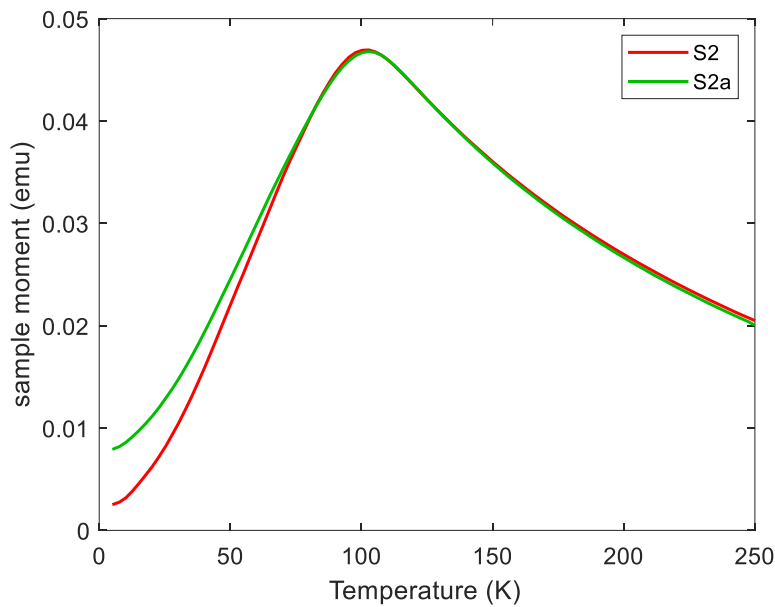

**Figure S2**. ZFC *dc* magnetization curves of the fresh- and aged-*S2* sample

Both curves were recorded after sample cooling in "zero" applied field from room temperature to 5 K.

The difference between the curves at low temperature is attributed to a difference in the residual field from the magnet of the SQUID between the two cooling processes. In both cases this residual field will have been small (less than 1 Oe) but it was probably larger in the case of the cooling of *S2a* than the cooling of *S2*. An applied field of 5 Oe was used for both measurements.

In contrast to *S1*, no aging-induced change in $T_{max}$ is found in *S2*, which is consistent with the observation that $M_S$ remains constant with aging in this sample (figure S1 of this supplementary Material).

## III. Oleic acid corrected magnetization values

The OA-corrected (mass) magnetization values were determined simply as

$$M = \frac{|\mathbf{m}|}{(1 - f_{OA})m}$$

where $|\mathbf{m}|$ denotes the magnitude of the sample magnetic moment, $f_{OA}$ is the OA mass fraction and $m$ is the "total" (OA + iron oxide) sample mass. The diamagnetic contribution to $|\mathbf{m}|$ due to the OA content of each sample was deemed to be negligible (using the highest value of diamagnetic susceptibility from published tables,[1] we estimate that the absolute size of this diamagnetic contribution at $H_{max}$ = 50 kOe is 4 orders of magnitude smaller than the sample moment detected at that field in each hysteresis loop of figure 1 of the main article).

---

[1] Broersma S 1949 *J. Chem. Phys.* **17** 873

# IV. M(H) data from S1 and *S1a* at different temperatures

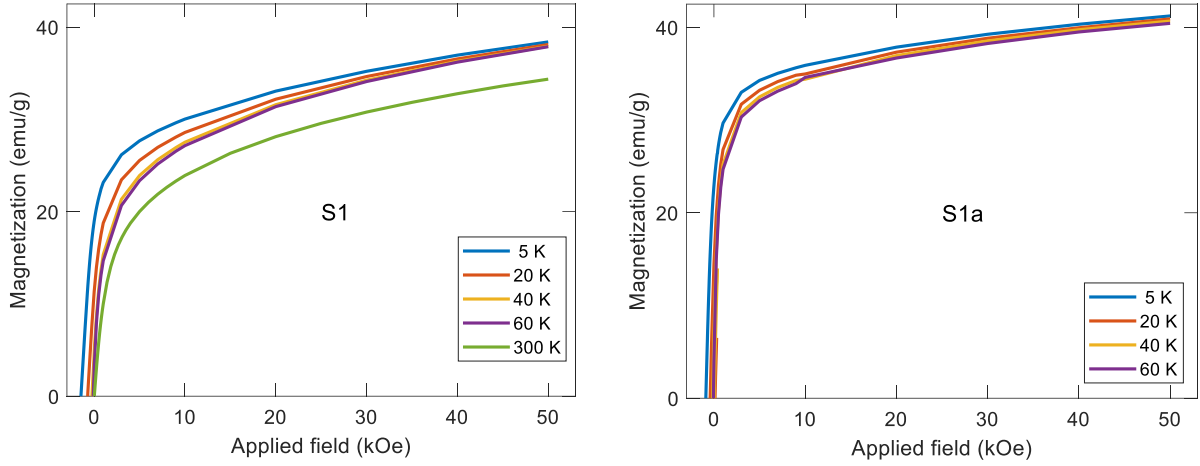

**Figure S3. M(H) dependence at different temperatures in fresh- and aged-*S1***
Left panel: *S1* (fresh sample); right panel: *S1a* (aged sample). Each curve was recorded upon decreasing field from 50 kOe, after an initial field cooling from ambient temperature to 5 K in H = 50 kOe (except, obviously, for the case of the 300 K data for *S1* – left panel).

# V. Analysis of widths of XRD peaks: determination of grain size of NP cores

Each XRD pattern shown in figure 5 of the main article contains five Bragg reflections corresponding to a spinel ferrite structure. With increasing scattering angle ($2\theta$), the reflections correspond to (*hkl*) of (220), (311), (400), (422) and (511). The fit to each pattern is based on a profile-matching ("Le Bail") technique, using a gaussian profile for each reflection, whose height and "width" (full width at half maximum, FWHM) are free (fitting) parameters. The positions (centers, $2\theta_{hkl}$) of the five gaussians are constrained by a single (cubic) lattice parameter, $a$, namely $2\theta_{hkl} = 2\sin^{-1}(\lambda/[2d_{hkl}]) = 2\sin^{-1}(\lambda\sqrt{h^2+k^2+l^2}/[2a])$, where $\lambda$ is the x-ray wavelength. The value of $a$ is a fitting parameter for each pattern. The data were background corrected before fitting. Deconvolution of each pattern with the resolution function of the diffractometer has a negligible effect on the fitted widths of the reflections.

In the figures shown below we investigate the applicability of a simple procedure – the Williamson-Hall (W-H) method[2] – to separate the microstrain and size (Scherrer) contributions to the reflection widths. The W-H formula is the following, where $\beta$ is the FWHM expressed in radians:

$$\beta \cos\theta = C\varepsilon \sin\theta + \frac{K\lambda}{D_{Scherrer}}$$

i.e., the formula is of the linear form:    $y \ = A \quad x \ + \quad B$

---

[2] Williamson G K and Hall W H 1953 *Acta. Metall.* **1** 22

Hence, via linear regression of the points in a W-H plot, one may, *in principle*, determine the microstrain parameter ($C\varepsilon$) and grain size ($D_{Scherrer}$). *The method, however, assumes isotropic microstrain.*

On the right we show the W-H plots based on the fitted reflection widths obtained from each pattern in figure 5 of the main article (the product of reflection width $\beta$ and the cosine of the Bragg angle, $\theta_{hkl}$, is plotted on vertical axis in each plot): the upper panel (blue circles) is the plot for *S1a*; the lower panel (red triangles) is the plot for *S2a*.

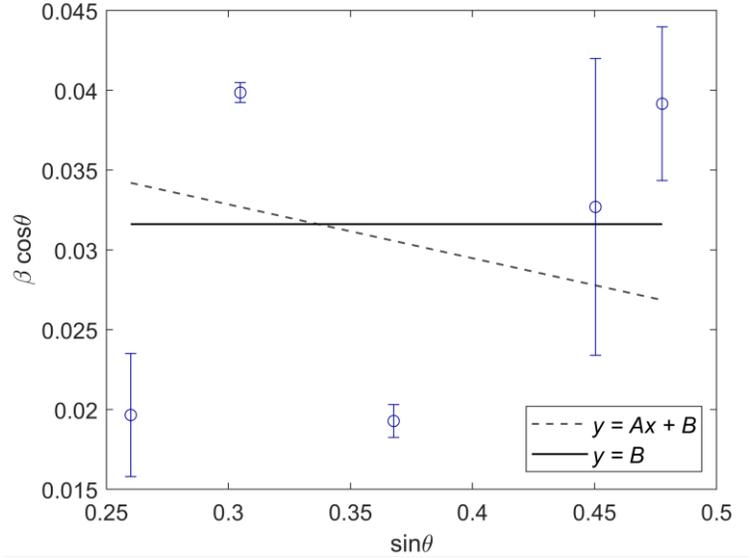

Neither set of points in the W-H plots forms a linear correlation. Therefore, we conclude that the W-H method is not an appropriate procedure for separating the strain and size contributions to the XRD from these NPs. The application of a more sophisticated technique (involving, e.g., the inclusion of anisotropic strain contributions[3]), beyond our present scope, would be required to separate these contributions.

We may use the plots, however, to make an estimate of grain size, by assuming that the Scherrer contribution to the widths of the Bragg reflections dominates the contribution due to microstrain.

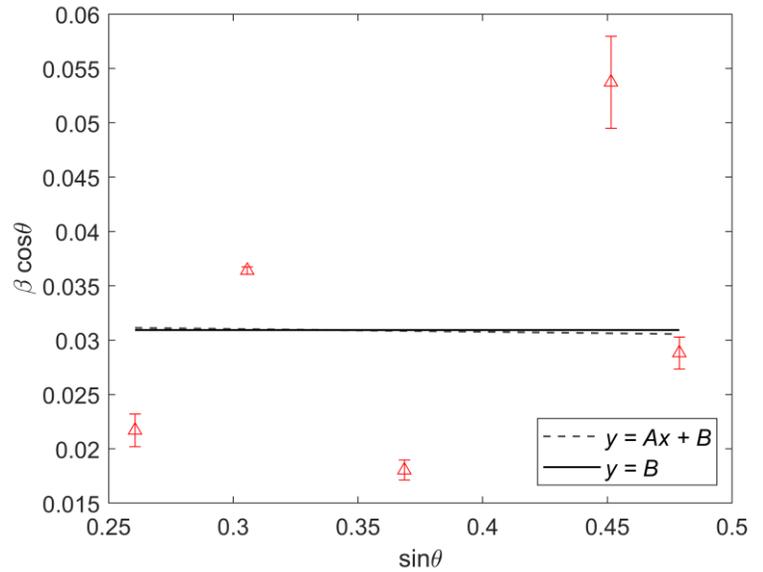

Using a weighted average of the $\beta \cos \theta$ values of each plot (equivalent to each fit of type $y = B$) and a value of $K = 4/3$ (appropriate for spherical particles[4]), we obtain $D_{Scherrer}$ values of:

- (6.5 ± 1.0) nm for *S1a*
- (6.64 ± 0.96) nm for *S2a*

For the sake of comparison, equating the $\beta \cos \theta$ value of the strongest reflection – the (311) – of each XRD pattern to $\frac{K\lambda}{D_{Scherrer}}$, we obtain:

- (5.15 ± 0.08) nm for *S1a*
- (5.64 ± 0.05) nm for *S2a*

---

[3] Muhammed Shafi P and Chandra Bose A 2015 *AIP Adv.* **5** 057127

[4] Chen D X *et al.* 2009 *J. Appl. Phys.* **105** 083924

## VI. Estimation of the NP core superspin

The extraction of the $M_{cores,i}$ values (where the subscript $i$ denotes *S1*, *S1a* or *S2*) is indicated in the figure shown on the right (the figure presents the same data as figure 2 of the main article). Figure S1 of this supplementary material implies that $M_{cores,S2a} \approx M_{cores,S2}$.

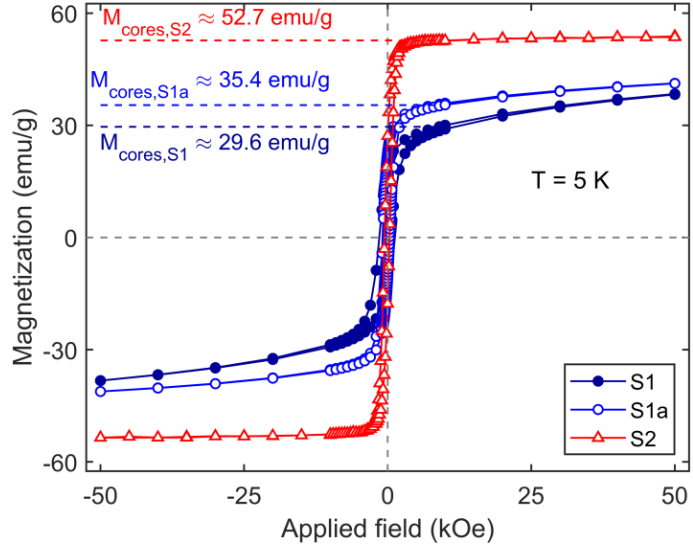

We may express each superspin, $\mu_i$, as the magnetic moment due to all the ferrimagnetic NP core moments (aligned by a field that is not large enough to achieve any significant alignment of spins in the disordered shells), $|\mathbf{m}_{cores,i}|$, divided by the total number of NPs in the given sample, $N_i$,

$$\mu_i = \frac{|\mathbf{m}_{cores,i}|}{N_i} = \frac{|\mathbf{m}_{cores,i}|}{(1-f_{OA,i})m_i/m_{NP,i}} = \frac{|\mathbf{m}_{cores,i}|}{(1-f_{OA,i})m_i}\rho_i V_{NP,i} = M_{cores,i}\,\rho_i\,\frac{\pi D_{TEM,i}^3}{6}$$

where $N_i$ has been expressed as the oleic acid-corrected assembly mass (see section III of this supplementary material) divided by the mean individual NP mass, $m_{NP,i} = \rho_i V_{NP,i}$, where $\rho_i$ is the NP (mass) density and $V_{NP,i}$ is the NP volume. Each $M_{cores}$ value (defined in section 3.1 of the main text and in the figure above) corresponds to $|\mathbf{m}_{cores,i}|$ divided by the *total mass of iron oxide material* (which obviously includes the mass of the spin-disordered shells). $D_{TEM,i}$ denotes the mean NP diameter (obtained by TEM).

## VII. Estimation of growth of the core radius with aging in *S1*

Our intention here is to estimate the (possible) variation with aging in the radius of the ferrimagnetically ordered core ($r_{core}$) of the *S1* NPs. Denoting the *volume magnetization* of the ferrimagnetically ordered NP core[5] as $M_{V,core}$ ($= \mu_i/V_{core,i}$), the core volume as $V_{core,i}$, and the ratio $M_{V,core,S1}/M_{V,core,S1a}$ as $R$, we may write (employing the final expression for $\mu_i$ given in the formula of the previous section)

$$\frac{\mu_{S1}}{\mu_{S1a}} = R\frac{V_{core,S1}}{V_{core,S1a}} = R\left(\frac{r_{core,S1}}{r_{core,S1a}}\right)^3 = \left(\frac{M_{cores,S1}}{M_{cores,S1a}}\right)\left(\frac{\rho_{S1}}{\rho_{S1a}}\right)\left(\frac{D_{TEM,S1}}{D_{TEM,S1a}}\right)^3$$

---

[5] Not to be confused with the magnetization as measured by a bulk technique like SQUID magnetometry. A microscopic technique such as magnetic neutron diffraction would probe this magnetization.

and therefore

$$r_{core,S1} = \frac{1}{R^{1/3}} \left(\frac{M_{cores,S1}}{M_{cores,S1a}}\right)^{1/3} \left(\frac{\rho_{S1}}{\rho_{S1a}}\right)^{1/3} \left(\frac{D_{TEM,S1}}{D_{TEM,S1a}}\right) r_{core,S1a}$$

The variation with aging in the radius of the core is then

$$\Delta r_{core} = r_{core,S1a} - r_{core,S1} = \left\{1 - \frac{1}{R^{1/3}} \left(\frac{M_{cores,S1}}{M_{cores,S1a}}\right)^{1/3} \left(\frac{\rho_{S1}}{\rho_{S1a}}\right)^{1/3} \left(\frac{D_{TEM,S1}}{D_{TEM,S1a}}\right)\right\} r_{core,S1a}$$

In the table below, we present four permutations in the calculation of $\Delta r_{core}$. Namely, we evaluate the above expression using $R$ equal to a typical ratio of magnetization values of bulk magnetite to bulk maghemite, and also for $R = 1$ (i.e. assuming $M_{V,core,S1}$ is unchanged with aging), and, for each $R$, we use values of $\rho_{S1}$ and $\rho_{S1a}$ of 5.2 and 4.9 g/cm³ (the bulk values of magnetite and maghemite), respectively, as well as evaluating for the case in which the density is unchanged with aging. We also assume that the geometrical ("TEM") particle size of $S1a$ would be indistinguishable[6] from that of $S1$ (i.e. $D_{TEM,S1}/D_{TEM,S1a} \approx 1$), and we use a value of $r_{core,S1a} = D_{Scherrer;S1a}/2 = 3.25$ nm (as obtained by XRD).

| $R = \dfrac{M_{V,core,S1}}{M_{V,core,S1a}}$ (estimated values) | $\left(\dfrac{M_{cores,S1}}{M_{cores,S1a}}\right)^{1/3}$ | $\left(\dfrac{\rho_{S1}}{\rho_{S1a}}\right)^{1/3}$ | $\Delta r_{core}$ (nm) |
|---|---|---|---|
| 1.16 | $\left(\dfrac{29.6}{35.4}\right)^{1/3} = 0.94$ | Different densities assumed: $\left(\dfrac{5.2}{4.9}\right)^{1/3} = 1.02$ | 0.28 |
|  |  | Same densities assumed: $\left(\dfrac{\rho_{S1}}{\rho_{S1a}}\right)^{1/3} = 1.00$ | 0.34 |
| 1.0 |  | Different densities assumed: $\left(\dfrac{5.2}{4.9}\right)^{1/3} = 1.02$ | 0.13 |
|  |  | Same densities assumed: $\left(\dfrac{\rho_{S1}}{\rho_{S1a}}\right)^{1/3} = 1.00$ | 0.20 |

---

[6] It is possible that the long-term ambient oxidation of $S1$ NPs leads to a slight NP swelling, i.e. as additional oxygen ions become incorporated into each NP volume, a slight enhancement of the geometric particle diameter occurs. Due to the experimental difficulties inherent in dispersing in solution the agglomerated NPs of a dried NP powder sample (which would be required in order to prepare monolayers of the aged-NPs on copper grids for size analysis by TEM), this possibility has not been investigated in the main article. Incidentally, if such swelling did occur then the $D_{Scherrer}/D_{TEM}$ ratio of $S1a$ would be smaller still than the value in table 1 of the main article, implying a relatively thicker crystallographically-disordered shells.

# VIII. Estimation of ratios of $E_{dd}$ parameters

From the relationship $E_{dd} \propto \mu^2/d^3$, the ratio of $E_{dd}$ parameters of *S1* to *S2* or of *S1* to *S1a* is given by[7]

$$\frac{E_{dd;S1}}{E_{dd;j}} = \frac{(\mu^2/d^3)_{S1}}{(\mu^2/d^3)_j} = \left(\frac{\mu_{S1}}{\mu_j}\right)^2 \left(\frac{d_j}{d_{S1}}\right)^3 \qquad (1)$$

where the subscript *j* denotes *S2* or *S1a*. We assume that the following relationship holds for the NP-NP separation in each assembly: 
$$d_i \propto \frac{D_{TEM,i}}{\phi_i^{1/3}}$$

where $\phi_i$ is the magnetic NP particle packing fraction. Substituting this relationship together with the final expression for $\mu_i$ (in section VI of this supplementary material) into equation (1),

$$\frac{E_{dd;S1}}{E_{dd;j}} = \left(\frac{M_{cores,S1}}{M_{cores,j}}\right)^2 \left(\frac{\rho_{S1}}{\rho_j}\right)^2 \left(\frac{D_{TEM,S1}}{D_{TEM,j}}\right)^3 \left(\frac{\phi_{S1}}{\phi_j}\right)$$

In the table below, each $E_{dd}$ ratio (*S1:S2* and *S1:S1a*) is evaluated in two ways: (i) by assuming $\rho_{S1}$ and $\rho_j$ as being the bulk values of magnetite and maghemite (5.2 and 4.9 g/cm³), respectively, and (ii) by assuming that *S1* and *S2* NPs have essentially the same density and that the density is essentially unchanged with aging (i.e. from *S1* to *S1a*). Values of $D_{TEM,S1}$ and $D_{TEM,S2}$ from the main article (8.6 and 8.0 nm, respectively) are used together with estimated values of $\phi_{S1}$ and $\phi_{S2}$ of 0.55 and 0.50, respectively.[8] We also assume (again) that the geometrical ("TEM") particle size of *S1a* would be indistinguishable from that of *S1* (i.e. $D_{TEM,S1a} = D_{TEM,S1}$), and that magnetic NP particle packing fraction is unchanged from *S1* to *S1a*.

| *j* | $\left(\frac{M_{cores,S1}}{M_{cores,j}}\right)^2$ | $\left(\frac{\rho_{S1}}{\rho_j}\right)^2$ | | $\frac{E_{dd;S1}}{E_{dd;j}}$ |
|---|---|---|---|---|
| **S2** | $\left(\frac{29.6}{52.7}\right)^2 = 0.32$ | Different densities assumed: $\left(\frac{5.2}{4.9}\right)^2 = 1.13$ | | 0.49 |
| | | Same densities assumed: $\left(\frac{\rho_{S1}}{\rho_j}\right)^2 = 1$ | | *0.43* |
| **S1a** | $\left(\frac{29.6}{35.4}\right)^2 = 0.70$ | Different densities assumed: $\left(\frac{5.2}{4.9}\right)^2 = 1.13$ | | 0.79 |
| | | Same densities assumed: $\left(\frac{\rho_{S1}}{\rho_j}\right)^2 = 1.00$ | | *0.70* |

---

[7] $\mu$ is the "average" magnitude of superspin of the ferrimagnetically NP cores. This superspin will not be exactly constant over the temperature range of each ZFC curve (the magnetization of the cores will have a weak temperature dependence, reducing by around 10 % from 5 K to 300 K). However, since equation (1) contains a ratio of $\mu$ values, we are justified in using magnetization data recorded at a single temperature (T = 5 K) to evaluate this expression.

[8] The value of $\phi_1 = 0.55$ has been estimated bearing in mind that $\phi = 0.59$ is found in assemblies of similarly sized NPs in which the OA content was only 3 % (after most of it had been removed by repeated washing) – see Normile *et al*. 2016 *Appl. Phys. Lett.* **109** 152404.

# IX. Assessment of interparticle interaction strength in *S1* and *S2*

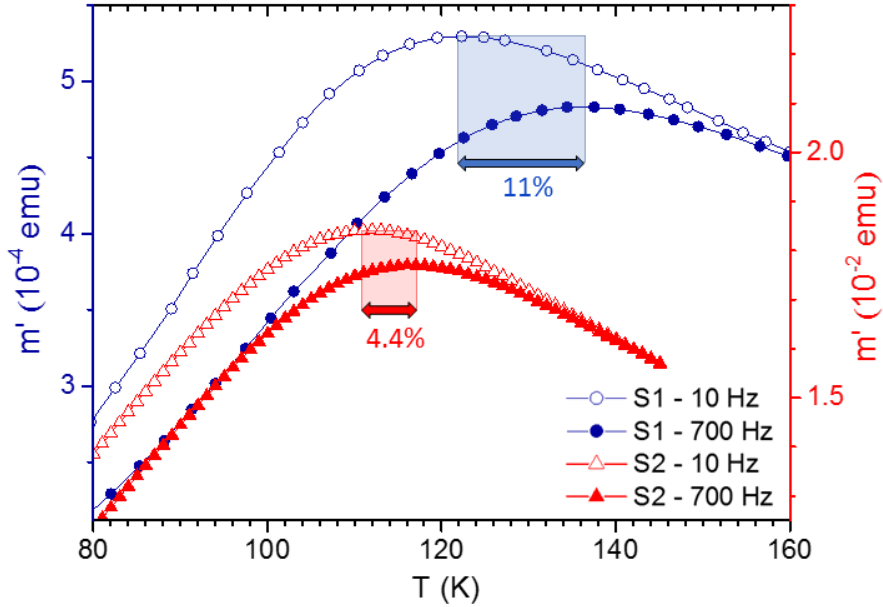

**Figure S4. In-phase *ac* magnetization at two different frequencies in *S1* and *S2***
The temperature dependences were measured at driving field of frequencies of 10 Hz and 700 Hz, and field amplitude 2.5 Oe. The circles refer to the left (vertical) axis and the triangles refer to the right axis. The relative extent to which the peak position ($T_{peak}$) shifts with a given change in measurement frequency ($f$) is related to the normalized strength of the interparticle dipolar interactions, i.e. to the size of the $E_{dd}/(K_{eff}V_{NP})$ ratio (defined in the main article); the larger this ratio, the smaller the *relative* shift.[9] The shift is considerably more pronounced in the case of *S1*, indicating a smaller $E_{dd}/(K_{eff}V_{NP})$ ratio in that sample relative to *S2*.

In the literature on strongly interacting magnetic particle systems, the size of the shift is quantified by the *total frequency shift parameter* ($p$), defined[10] as $p = \dfrac{\Delta T_{peak}/\langle T_{peak}\rangle}{\Delta \log_{10}(2\pi f)}$, where $\Delta T_{peak}$ is the absolute peak shift, $\langle T_{peak}\rangle$ is the average peak position (for the two $f$ values) and the ratio $\Delta T_{peak}/\langle T_{peak}\rangle$ is the *relative* peak shift. From the data in the figure above, we obtain $p$ = 0.060 for *S1* and $p$ = 0.024 for *S2*. The $p$ value for *S1* implies only moderate interactions.

---

[9] For a system with a small $E_{dd}/(K_{eff}V_{NP})$ ratio, NP blocking will be governed by the individual NP anisotropy energy and the system dynamics will be Arrhenius-like; the shifts in $T_{peak}$ upon variation in the $f$ will be relatively large. As the value of the ratio increases – by, for example, increasing $E_{dd}$ or reducing $K_{eff}V$ – the dynamics will eventually depart from an Arrhenius-like regime and pass into a *critical slowing down* regime (characteristic of a superspin glass), where the relative shift in $T_{peak}$ due to variation in $f$ is reduced relative to the Arrhenius regime.

[10] See, for example, De Toro *et al.* 2013 *Appl. Phys. Lett.* **102** 183104

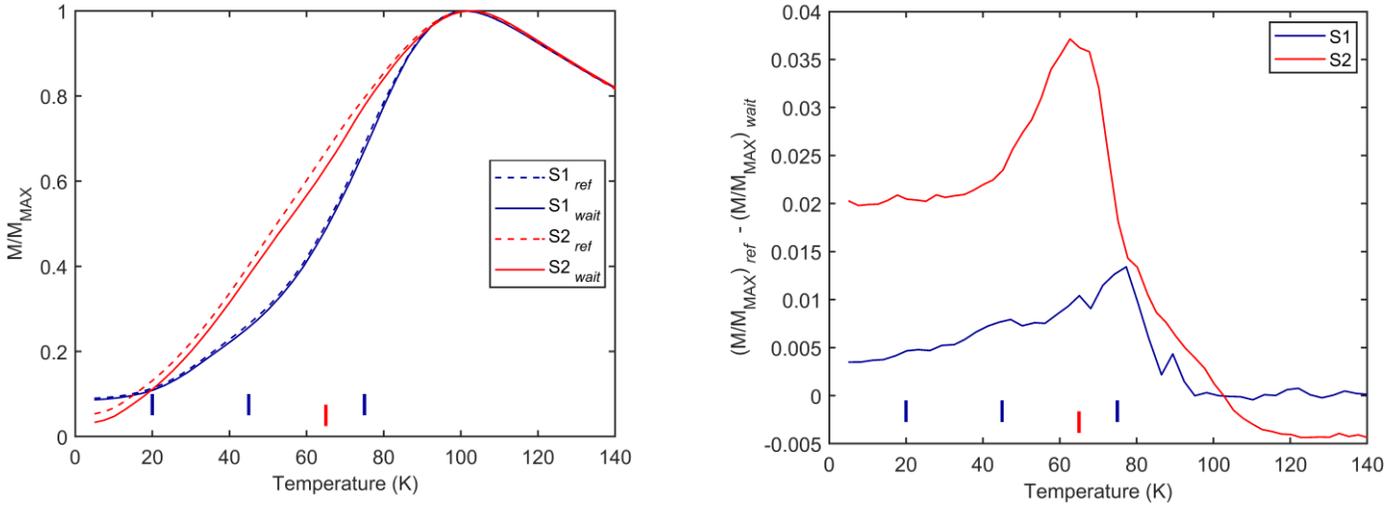

**Figure S5. ZFC memory measurements on *S1* and *S2***
Left panel: low-field (H = 5 Oe) temperature-dependent ZFC *dc* magnetization curves performed on each sample after cooling directly to 5 K (dashed-line curves labelled "*ref*") and after temporarily halting the cooling process at the temperature(s) indicated (for each sample) by the short vertical bar(s), for a waiting time of 3 hours per halt (solid-line curves labelled "*wait*"). Three successive halts were made during the cooling of *S1* (at 75, 45 and 20 K) and a single halt (at 65 K) was performed in the case of *S2*.

Right panel: the memory effect ("*ref – wait*" difference curve) of each sample, obtained as $\left[\left(\frac{M}{M_{MAX}}\right)_{ref} - \left(\frac{M}{M_{MAX}}\right)_{wait}\right]$, where $M_{MAX}$ denotes the magnetization registered at the peak of each ZFC curve.

A well-defined ZFC memory effect is characteristic of a superspin glass (SSG) phase formed by a dense assembly of magnetic (oxide) NPs.[11] Of the two samples in the right panel, only *S2* shows a well-defined memory effect, which is consistent with its lower *p* value relative to *S1*. We suggest that the *S1* sample is a close-to-borderline SSG system.

The non-zero difference signal below around 40 K and above around 110 K in the *S2* memory effect curve is attributed to a difference in the residual field from the magnet of the SQUID between the two corresponding cooling processes ("*ref*" and "*wait*") performed on this sample.

---

[11] See, for example, De Toro *et al*. 2013 *Appl. Phys. Lett.* **102** 183104

# X. Note on DMF factors and the effect of reducing interparticle interactions between the magnetic NP cores of *S2* on the $\chi''$ line-shape

Note: The $\chi''$ curve of *S1* will also be affected – broadened – by DMF effects, however, our understanding is that these effects do not lead to an asymmetric shape of the main peak because the underlying (intrinsic) line-shape is symmetrical. Indeed, the degree of DMF effects may very well be similar between *S1* and *S2* since a higher ϕ is expected in *S1* relative to the *S2*, which will give rise to a higher DMF factor, however DMF effects also depend on the value of the (*external*) sample magnetization,[12] which will be lower in *S1*.

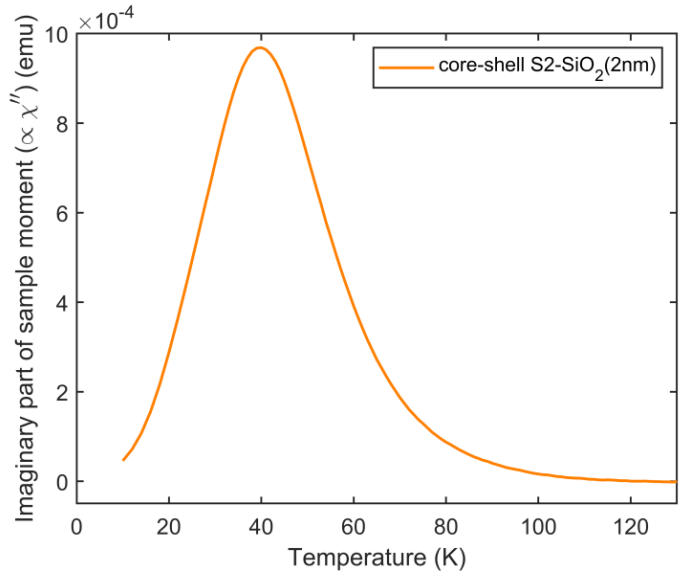

**Figure S6. Out-of-phase *ac* susceptibility in core-shell *S2*-SiO$_2$(2nm) NPs**

Temperature dependence of the out-of-phase component of the sample magnetic moment (proportional to $\chi''$) from *ac* magnetometry, measured with a driving field of frequency 10 Hz and amplitude 1 Oe in a pressed disk sample composed of iron oxide NPs from the same synthesis batch as *S2* but with the oleic-acid shells substituted by 2 nm thick silica shells. (Details on the silica coating technique can be found in a previous study[13] where this same sample was investigated.)

The silica shells cause an increase in the separation between nearest-neighbour magnetic NP cores with respect to the *S2* sample, leading to a reduction in the strength of interparticle magnetic dipolar interactions relative to the interactions in that sample, which, in turn, causes a reduction in temperature of the peak position in the $\chi''(T)$ curve as well as a loss of the intrinsic asymmetrical line-shape that resulted from the strong interparticle interactions in *S2*. The increased core-core separation also implies a reduced magnetic NP packing fraction with respect to *S2*, which gives rise to a reduction in the peak broadening effects due to demagnetizing fields[12,14] compared to the case of *S2*. No hump feature that could be associated with the SSF transition at the iron oxide particle surfaces is observed near to the EB onset temperature of *S2* (T$_E$ = 30 K, which is identical to EB onset temperature found for these silica-coated NPs) in the above $\chi''(T)$ curve.

---

[12] Normile *et al*. 2016 *Appl. Phys. Lett.* **109** 152404

[13] De Toro *et al*. 2017 *Chem. Mater.* **29** 8258

[14] The effects are reduced due to a reduction in both the DMF factor and the (*external*) sample magnetization, both relative to the *S1* sample.